\documentclass[10pt,twocolumn,conference]{IEEEtran}
\usepackage{comment}
\usepackage{subfigure}
\usepackage{color}
\usepackage{epsfig}
\usepackage{pslatex}
\usepackage{caption}
\usepackage{amssymb, amsmath}
\usepackage{epsfig}
\usepackage{algorithm}
\usepackage{algorithmic}
\usepackage{multirow}
\usepackage{flushend}
\usepackage{array}
\usepackage{url}
\usepackage{multicol}
\usepackage{lastpage}
\usepackage{fancyhdr}
\usepackage{balance}

\newenvironment{itemizer}{\begin{itemize}
                \setlength{\parsep}{-.2cm}
                \setlength{\itemsep}{-.0em}}{\end{itemize}}

\graphicspath{{./graphics}}
\begin{document}

\title{Enhancing MapReduce Fault Recovery Through Binocular Speculation}

\author{
	\begin{tabular}{ccccc}
		Huansong~Fu & Yue~Zhu & Amit~Kumar~Nath & Md.~Muhib~Khan & ~Weikuan~Yu
	\end{tabular} \\
	\hspace{1pc}~\\
	\begin{tabular}{c}
		Department of Computer Science, Florida State University \\
		{\{fu,yzhu,nath,khan,yuw\}@cs.fsu.edu}
	\end{tabular}
}

\maketitle

\begin{abstract}
MapReduce speculation plays an important role in finding potential task stragglers and failures. But a tacit dichotomy exists in MapReduce due to its inherent two-phase (map and reduce) management scheme in which map tasks and reduce tasks have distinctly different execution behaviors, yet reduce tasks are dependent on the results of map tasks. We reveal that speculation policies for fault handling in MapReduce do not recognize this dichotomy between map and reduce tasks, which leads to an issue of speculation myopia for MapReduce fault recovery. These issues cause significant performance degradation upon network and node failures. To address the speculation myopia caused by MapReduce dichotomy, 
we introduce a new scheme called binocular speculation to help MapReduce 
increase its assessment scope for speculation. 
As part of the scheme, we also design three component techniques including
neighborhood glance, collective speculation and speculative rollback. 
Our evaluation shows that, with these techniques, binocular speculation 
can increase the coordination of map and reduce phases, and enhance the efficiency of MapReduce fault recovery. 

\end{abstract}


\section{Introduction} 
\label{sec:intro} 
MapReduce has gained wide popularity since Google 
introduced it~\cite{osdi/googlemr04} in 2004.
Specifically, Hadoop and its successor YARN~\cite{yarn_url} 
are popular open-source implementations of MapReduce.
Many previous works have focused on improving the performance and
scalability of MapReduce~\cite{osdi/googlemr04,sigmod:hadoop:dbms}, 
task and job scheduling~\cite{delay:eurosys10, wang:icac13},
data redundancy and availability etc~\cite{raft:icde2011}. However, given the growing scale of hardware, firmware, and system components leveraged in
computer systems, the mean time between
failures or interruptions (MTBF/I) will be around 6.5-40
hours~\cite{3-wang2010hybrid, 4-kharbascombining}.
It is expected that the resilience challenges will be further
compounded by the advances in system technologies.
Therefore, besides the paramount interest on efficient analytics,
it is important to investigate the fault resilience of MapReduce
and its impact on data analytics.

MapReduce adopts a two-phase (map and reduce) scheme
to support many user applications (large batch jobs and small interactive queries),
and schedule their tasks.  To maximize the use of system capacity and
ensure fairness among different jobs,
existing task scheduling policies
have focused on fair distribution of resources among
tasks~\cite{delay:eurosys10, DRFsched}. 
The execution behaviors of these two phases in MapReduce are distinctly different.
In the map phase, tasks are typically short-lived and are launched repetitively to
balance the use of available map resources.  Tasks in the reduce phase are
typically longer in duration and start after the completion of the first map task. 
To compound the situation, there is a dependence between the two phases, i.e.,
reduce tasks must fetch intermediate data generated by map tasks.
These two distinct system execution behaviors lead to a fundamental {\bf dichotomy} 
between the two phases of MapReduce.

For fault resilience,
MapReduce adopts a simple speculative
task re-execution mechanism to launch redundant tasks.
All tasks are treated in a similar manner for resource allocation, 
process speculation, synchronization and re-execution.
We have revealed that the fundamental dichotomy of
MapReduce can cause a number of efficiency and resilience issues.
Particularly, it causes different requirements
from map and reduce tasks in terms of resource allocation, speculation
policies, and task temporal relationships.  
Such distinct requirements on resources can lead to the idleness of reduce tasks in big jobs
and delay the completion of small jobs~\cite{jian:sigmetrics12,wang:icac13,delay:eurosys10}.

In addition, the dichotomy can lead to shortsighted decisions for speculative
execution in MapReduce, a phenomenon we refer to as {\em Speculation Myopia}.
It prevents a newly launched task, e.g., a reduce task from recognizing its data
dependence on the intermediate data, which might have been damaged or lost on
another node. Myopic speculation can only detect the need for launching
additional tasks until multiple attempts fail.  Furthermore, the dichotomy 
leads to serious fault handling issues such as wasteful redundant execution 
and degradation of system efficiency (See further details in Section~\ref{sec:motivation}).

To this end, we have conducted an analysis on MapReduce
speculation and characterized the impact of 
{\em Speculation Myopia} on MapReduce fault recovery.
We have characterized two main symptoms of speculation myopia:
{\em dependency-oblivious speculation} and {\em scope-limited speculation}.
Accordingly, we have designed a new scheme called binocular speculation to 
increase the scope of MapReduce speculation with three component techniques: 
{\em neighborhood glance}, {\em collective speculation} and {\em speculative roll-back}.
Our experimental results show that binocular speculation not only heals the speculation
myopia while handling the failure-related stragglers,
but also improves performance and scalability under heavy workloads.
To the best of our knowledge, binocular speculation is the first to examine 
the failure-related stragglers in MapReduce with an effective mitigation method.

In summary, our paper makes the following contributions:
\begin{itemizer}
	\item We have analyzed MapReduce speculation, revealed the existence
	of speculation myopia, and further characterized its causes and effects.
	\item We have introduced a new scheme called {\em Binocular Speculation} to heal the
	impact of speculation myopia, along with three techniques: neighborhood
	glance, collective speculation, and speculative roll-back.
	\item We have conducted an extensive set of experiments to evaluate binocular
	speculation. The results demonstrate that our new speculation scheme
	can heal all symptoms of speculation myopia for MapReduce fault recovery.
 
\end{itemizer}

\section{Background and Motivation}
\label{sec:motivation}
\subsection{Overview of YARN MapReduce}
\label{subsec:yarn}
As a resource management infrastructure, YARN aims to simultaneously 
support various programming models, such as MapReduce~\cite{osdi/googlemr04}.
  We focus on the YARN MapReduce programs in this project.
The execution of MapReduce programs 
includes two major phases: {\em map} and {\em reduce}.  
YARN supports such execution through a ResourceManager and several NodeManagers. 
The ResourceManager manages all resources 
and allocates containers to running applications.
Each NodeManager abstracts the resources on the node as multiple {\em containers}
to serve the need of different applications. 

Each YARN job starts with one ApplicationMaster {\em a.k.a MRAppMaster},
which negotiates with the ResourceManager for containers.
When granted, it launches map tasks. Each map task applies the map
function to an input split of many $<$k,v$>$ pairs and generates intermediate
data that are organized as a Map Output File (MOF). 
When MOFs are available,  
the MRAppMaster launches reduce tasks,
overlapping the reduce phase with the map phase of remaining map tasks. 
A reduce task is a combination of two 
stages: {\em shuffle/merge} and {\em reduce}. In the former stage,
it fetches and merges its partitions from all MOFs.  
It then enters into the reduce stage, where the reduce function is applied to
the intermediate data. 
The final results are stored to the Hadoop Distributed File System (HDFS).

\subsection{The Dichotomy of MapReduce}
MapReduce offers a simple programming model for large scale
analytics applications. 
Its map and reduce phases have very distinct execution behaviors,
causing a variety of disparities in task management, 
resource allocation and fault handling.
We refer to these disparities as {\em the Dichotomy of MapReduce}.
This dichotomy leads to a disparity in the scheduling and resource
management of map and reduce tasks~\cite{jian:sigmetrics12,wang:icac13,tan:sigmetrics14}.
The phases of Map tasks are typically short-lived. Their execution model 
can be represented as a typical processor sharing queue.
Reduce tasks 
shuffle and process the intermediate data generated by more map tasks. 
Their execution model is a typical multi-server queue.
To compound the situation, these two queues are dependent through a constraint that 
the reduce tasks in the multi-server queue must fetch intermediate data generated by 
the map tasks from the processor sharing queue. 
For Short-lived map tasks, it is easy to balance among available containers for processing data 
splits, but long-running reduce tasks can retain their containers while 
waiting on map tasks for intermediate data.
 
Such distinct requirements on resources can lead to the idleness of reduce tasks in big jobs 
and delay the completion of small jobs~\cite{jian:sigmetrics12,wang:icac13,delay:eurosys10}.

In addition, the dichotomy can lead to shortsighted decisions for speculative
execution in MapReduce, a phenomenon we refer to as {\em Speculation Myopia}.
Speculative execution is an essential mechanism for MapReduce to deal with
stragglers and task failures~\cite{late:osdi08, ganesh:clones, nsdi14grass,
	osdi:mantri}. 
It monitors the progress of tasks and launches a
redundant copy of slow or failed tasks. The dichotomy of MapReduce
prevents a newly launched task, e.g., a reduce task from recognizing its dependence on the intermediate data, which might have been damaged or lost on
another node. Such myopic speculation can only detect the need of launching
additional tasks until multiple attempts fail.
Finally, for task resilience, MapReduce recovers the work of any delayed or faulty
task using a redundant or re-executed task, but the dichotomy 
leads to a disparity in the fault recovery of tasks.
We have revealed serious fault handling issues such as wasteful redundant execution, 
asymmetric task recovery and failure amplifications, degrading system efficiency.
The impact of these issues will be discussed in detail in the rest of this
section.

\subsection{MapReduce Speculation and Its Myopia}
\label{subsec:faultamp}

Many previous studies~\cite{schroeder:dsn06, moody:sc10} of 
failure characteristics  
have revealed that a large portion of failures are transient, ranging 
from 31\%~\cite{moody:sc10} to 65.2\%~\cite{dong:sc09}. 
MapReduce adopts a simple strategy via task and data 
regeneration to handle transient failures.
Once a task is detected as failed, 

it re-launches another attempt of the same task,
repeating the work achieved previously.
This works very well for short-lived map tasks whose amount of work
is generally much smaller compared to the reduce tasks,
but it is not as effective for reduce tasks because of their 
long-running behaviors~\cite{delay:eurosys10,wang:icac13}. 
In addition, MapReduce employs a global speculator for speculatively 
launching tasks to guard against stragglers or failures,
 
A global speculator proactively makes another attempt for a task that
is lagging behind (a.k.a, stragglers) so that any attempt that finishes sooner 
will help the job to progress further.
Similar speculation is adopted by other representative parallel computing
paradigms such as Dryad~\cite{isard2007dryad}.

{\bf Speculation Myopia:}
the existing speculation scheme in MapReduce 
is unable to peek through the dichotomy of its map and reduce phases, 
resulting in the shortsightedness of speculative tasks, 
i.e., Speculation Myopia. 
For example, a redundant reduce task cannot help when the failure was caused
by the loss of intermediate data from prior map tasks.
As a result, the redundant task will fail again. Myopic speculation 
also manifests itself in a number of other ways. 
The global speculator adopts a serial scheme to
analyze the progress of tasks and launches speculative tasks in a sequential manner
with a fixed delay interval. This is intended for an important cause, i.e., 
limiting additional resource consumption. Nonetheless, 
such serial speculation with fixed delays is unable to meet the need of 
many speculative tasks when a MapReduce platform is affected by 
network congestion or resource contention. 
In addition, speculation relies on the significant variation of progress among tasks. 
When tasks from a job are located on a single node or a few nodes which are equally
affected by a system condition, the global speculator is unable to 
launch any speculative task for the job because there is not enough progress variation
among its tasks. 

\begin{figure}[tbh]
	\begin{center}
		\vspace{-0.5pc}
		
		{\includegraphics[width=0.36\textwidth]{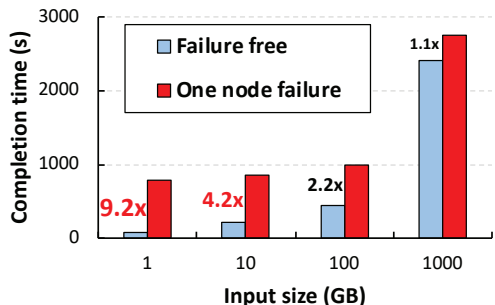}}
	
		\vspace{-0.5pc}
		\caption{Job Slowdown Caused by Node Failures}
		\vspace{-1.0pc}
		\label{fig:motiv-1g-1000g}
	\end{center}
\end{figure}

{\bf Impact to Fault Handling:}
We have found that the Speculation Myopia approach has some major issues for handling 
stragglers caused by various forms of system failures~\cite{osdi:mantri}.
Stragglers due to failures are not mitigated by the speculation mechanism,
especially those from small size jobs.
Fig.~\ref{fig:motiv-1g-1000g} shows the job slowdown caused by a single node failure,
which is extremely common in real-world MapReduce systems~\cite{dean2006experiences}.
When the job size is large, the slowdown is not obvious.
But when it comes to small jobs (1 GB to 10 GB), the slowdown 
can be as much as 4.6x to 9.2x to the normal job running time.
To make things worse, the majority of jobs on production systems are actually small size jobs.
It has been widely 
reported~\cite{chen2012interactive,ananthanarayanan2012pacman,ganesh:clones,appuswamy2013scale,reiss2012heterogeneity}
that the size of MapReduce jobs in production clusters follows the power-law distribution,
with a vast majority of them (e.g., more than 80\% for
Facebook workloads~\cite{ananthanarayanan2012pacman}) containing less than 10 GB input.

\subsection{Analysis of Speculation Myopia}
\label{sec:cause}

Failure is identified as one of the major causes of stragglers~\cite{osdi:mantri,dean2006experiences}.
We have closely examined the MapReduce speculation mechanism,
and found that, in many common failure scenarios, 
it can seriously impede the efficiency of straggler mitigation.
As mentioned before, this can lead to the shortsightedness of MapReduce
speculation, i.e., Speculation Myopia,  resulting in unsuccessful speculative decisions 
and wasteful system resources. For succinctness, we focus on 
two main symptoms for further elaboration.

\subsubsection{Dependency-Oblivious Speculation}
The existing speculation scheme in MapReduce 
is not very effective for reduce tasks because of their long-running
behaviors~\cite{delay:eurosys10,wang:icac13}.
It is unable to peek through the
dependency between its map and reduce phases, resulting in a symptom called
{\em Dependency-Oblivious Speculation}.
The speculative decision is made based on the progress of all running tasks. 
If a task is finished, it will be excluded from the candidates for further speculative execution. 

Intuitively, it is reasonable to consider only running tasks since completed tasks should have no 
way of delaying a job. However, MapReduce typically requires the 
use of intermediate data that is produced by the completed map tasks. 
If this intermediate data was lost, the job would be 
held up until it finally found out that the intermediate data was permanently lost. 
Furthermore, when a new attempt for the failed reduce task is launched,
the required intermediate data can still be missing.
Because of this, a redundant reduce task cannot help but wait and encounter several fetch failures again.
Thus, completed tasks could also become stragglers, which the current speculation 
mechanism is unable to address. 
To mitigate this issue, MapReduce needs to gain an awareness of task dependencies.

\subsubsection{Scope-Limited Speculation}

MapReduce speculation relies on the variation of progress among individual tasks.
Based on the progress reports from heartbeat messages, 
the global speculator of MapReduce
measures the progress variation among all tasks and selects one 
of the tasks that falls behind for speculative execution.
This works well when there is a sufficient number of tasks,
for example, when a big job is running on multiple nodes.
But when tasks from a job are located on a single node or a few nodes equally
affected by a system condition, the global speculator is unable to
launch any speculative task for the job because there is not enough progress variation
among its tasks.

For example, for small jobs such as those with 1 GB of input data, 
their tasks are very likely to be all situated on the same node. 
The failure or slowdown of that node will lead to the lack of progress reports from all of
them. Hence, the global speculator has no way of launching speculative tasks
until the timeout.
Clearly, the long delay should be avoided by early speculative tasks
as soon as MapReduce recognizes that the tasks on one node are all slow.
Thus the existing speculation mechanism
has a limited scope in measuring the variation of progresses. 
For early speculation decision, it needs to expand the scope of progress assessment 
from the cross-node comparison of a limited few tasks.
This way, MapReduce can analyze task progresses in a global manner and guard against
stragglers from an unhealthy segment of the system.

\section{Binocular Speculation for Failure Recovery}
\label{sec:design}
 We have explored the implementation of YARN speculator with
simple walk-around options. For example, we have
used a shorter timeout value before the speculator decides 
to launch a copy for a straggler task.
 Our unsuccessful attempts suggest that the symptoms 
of speculation myopia occur due to 
fundamental limitations of YARN speculation, and require an algorithmic 
renovation.
A complete solution requires a new speculation scheme.
 
\begin{figure}[htb]
	\begin{center}
		\vspace{-0.5pc}		
		{\includegraphics[width=0.48\textwidth]{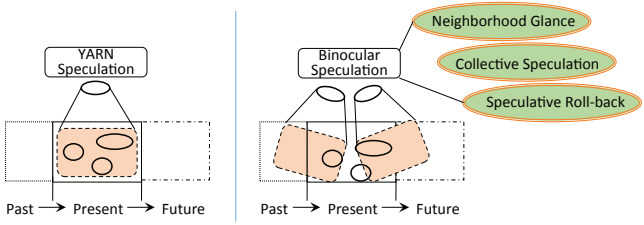}}
		\caption{Comparison of YARN and Binocular Speculation}
		\vspace{-0.5pc}
		\label{fig:binocular}
	\end{center}
\end{figure}

In this paper, we propose a new speculation scheme that 
can heal the vision of speculation with
better awareness of task dependencies and wider scope for measuring 
progress variations. 
We refer to the new scheme as {\em Binocular Speculation}.

As shown in Fig.~\ref{fig:binocular}, the current speculation in YARN MapReduce 
focuses entirely on the present state of task execution, without an awareness 
on task dependence with past and present tasks. 
In contrast, we introduce a new scheme that can expand
the scope of speculative decisions, 
taking into account the relationship of a task with other tasks in 
both directions of time: past and present, hence the name. 

Three new techniques are introduced in our proposed scheme.
We first design a {\em neighborhood glance} mechanism
to help the global speculator to increase the assessment scope of 
task progresses to the {\em neighboring} nodes and tasks, as well as
{\em glancing} over data and task dependencies across different phases.
Then we introduce a technique called {\em collective speculation} to 
allow a flexible frequency of task speculation so that speculative 
tasks can be launched dynamically, and collectively, if permitted by available 
resources. 
Finally, we develop {\em speculative rollback} to periodically log
the progress of tasks so that 
a speculative task can be launched by rolling back to a previous log 
and progress further, instead of starting from scratch.

\subsection{Neighborhood Glance}
\label{sec:neigh}

A MapReduce job involves many concurrent tasks across computer nodes. In the temporal dimension, the progress of tasks and nodes can vary from time to time, and in the spatial dimension, the progress of some tasks/nodes 
can also differ from that of other tasks/nodes.  
To better assess the progress variations in a MapReduce job, 
we introduce the {\em neighborhood glance} mechanism to measure
the progress variations and identify underperforming tasks/nodes 
around a spatial neighborhood for the local task/node, 
or a time range from the current time point.

In addition to progress assessment, 
we consider the responsiveness history of a node to
assess a node failure as well
for more swift recovery upon the node failure.
To find a slow or failed node, we take on three independent assessment policies.
Specifically, {\em spatial progress assessment} aims to find
a slow node that is significantly slower than its neighbors. 
{\em Temporal progress assessment} is intended to identify
a straggler that may happen to slow down compared to its progress history. 
Finally, {\em faulty node monitoring} is used to discover a disconnected node.
Next, we describe each type of assessment in more detail.

\subsubsection{Progress Assessment in the Spatial Neighborhood}
By default, YARN speculation measures the accumulative task progress score~\cite{late:osdi}
to determine the stragglers.
This metric can be easily affected by imbalanced task assignment or nonuniform workloads among jobs.
The notations in YARN denote node $N$ for job $J$ as $(N^J)$ and
$\rho(t_i)$ as the task progress rate of task $i$ that belongs to job $J$ on node $N$.
Furthermore, the progress is represented as $\rho$, where $\rho(t)=\frac{\zeta(t)}{\tau_{t}}$,
and $\zeta(t)$ is the {\ttfamily ProgressScore} of task $t$ and 
$\tau_{t}$ is the running time for $t$.

In neighborhood glance, we introduce a metric called {\ttfamily NodeProgressRate} ($P$)
to quantify the average task progress rate on a compute {\em node}.
For example, $P$ of $(N^J)$ is defined as $P(N^J)=Avg(\rho(t_i)_{{t_i}\in J})$,

To decide a slow node, we compare its $P$ with other nodes inside its neighborhood.
We denote $\sigma(P)$ as the standard deviation of $P$ within a neighborhood,
and $NH\{N_i\}$ as the collection of all nodes in $N_i$'s neighborhood.  
So if 
\begin{equation}
P(N^J) < avg(P(N_i^J)_{N_i \in NH\{N_i\}}) - \sigma(P(N_i^J)_{N_i \in NH\{N_i\}})
\end{equation}
we then mark node $N$ as a slow node for job $J$ in our speculation scheme.

\subsubsection{Progress Assessment in the Temporal Neighborhood}

To expand our assessment of progress into the historical temporal dimension,
 we introduce a metric called {\ttfamily NodeProgressChangeRate} ($\Delta$) to
keep track of the acceleration of {\ttfamily NodeProgress} $\zeta$:
\begin{equation}
\Delta(N^J)|_{T_i}=\frac{\zeta(N^J)|_{T_i} - \zeta(N^J)|_{T_{i-1}}}{T_i - T_{i-1}},
\end{equation}
where $\Delta(N^J)|_{T_i}$ is the {\ttfamily NodeProgressChangeRate} of $(N^J)$ at time $T_i$
and $\zeta(N^J)|_{T_i}$ is the summation of {\ttfamily ProgressScore} of all the tasks on $(N^J)$  at time $T_i$.

Note that we only consider the on-going tasks for the calculation of $\zeta$.
Hence, we can avoid aggressive speculation when many tasks have completed and
exited near the end, and the accumulative {\ttfamily ProgressScore} declines suddenly.
To deduce that $(N^J)$ is slow for job $J$ at time $T_i$, the following condition needs to be met:
\begin{equation}
\Delta(N^J)|_{T_i} < Threshold^{slowdown}  \times \Delta(N^J)|_{T_{i-1}}
\end{equation}
where $Threshold^{slowdown}$ is the slowdown threshold for $\delta$ 
that will determine a computer node as a straggler. This is a configurable parameter, by default $0.1$.

\subsubsection{Node Failure Assessment}
\label{sec:failure-assess}

Although a failed node can be detected by our task progress assessment, 
the speculator can take a while to detect its progress slowdown.
There are additional indicators of a node failure other than the execution progress,
e.g., the loss of NodeManager heartbeat.
YARN's ResourceManager receives a heartbeat of each NodeManager in every second.
A continuous stretch of lost heartbeats can affirm the node failure. 
But transient faults can cause lost heartbeats as well.

We use a heuristic algorithm to detect a node failure.
First of all, a node $N_i$ is marked as failed when the duration since its last response
exceeds a $Threshold^{fail}_i$. If a node is responsive in a heartbeat report, 
we check if it is a resuming heartbeat from a previously lost node. If that is the case, 
we measure the time duration that the node remains lost.
This time duration is used later for updating $Threshold^{fail}_i$.

To determine $Threshold^{fail}_i$, we use a window-based mechanism that can 
take the historical loss of responsiveness into account, where the earlier
loss of responsiveness has less impact on the threshold.
For instance, in order to capture the temporal locality between 
the last $L$ failures and the next failure at a node, 
we define the length of our window as $L$. We use $R_n$
to represent the duration of the node's lost responsiveness during the $n$-th window in $L$.
We hope to estimate the duration of lost responsiveness for the next failure
($P_{n+1}$) based on previous measurements.
Given any node $i$ and a window of $L$, $P_{n+1}$ can be estimated as follows:
\begin{equation}
P_{n+1} = \frac{\sum_{k=1}^{L}(2^{L+1-k} \times R_{n+1-k})}{\sum_{k=1}^{L}(2^k)} 
\end{equation}
The parameter $L$ is tuned based on the trade-off between the estimation accuracy 
and the computing overhead.

\subsection{Collective Speculation}

Once faults or failures are identified through neighborhood glance, 
we use a {\em collective speculation} mechanism to maximize the number of 
speculative tasks and minimize the progress delays caused by these faults or failures.
Such mechanism needs to balance the recovery speed and resource consumption.
Thus, instead of spawning speculative task attempts on all available compute nodes,
we start with the neighboring nodes. 
If there are enough containers in the neighborhood, 
all speculative tasks are launched.
If not, we use the available containers for those tasks.
After that, we need to gradually add speculative tasks.
To that purpose, we start with a small number of speculative tasks that equals to {\ttfamily COLL\_INIT\_NUM}
on compute nodes beyond the local neighborhood. 
We monitor the progress of both the speculative copy of a task and the original one.
If either task is completed, we terminate the other one.
If the speculative task has shown faster progress rate than the original copy,
we continue to launch more speculative tasks (with a multiplication 
factor {\ttfamily COLL\_MULTIPLY}) until all stragglers have had an speculative attempt.
In the $i$'s times of speculation, the number of speculative tasks to launch equals to
 {\ttfamily COLL\_INIT\_NUM} $\times$ {\ttfamily COLL\_MULTIPLY}$^i$.

\begin{figure}[htb]
	\begin{center}
		\vspace{-0.5pc}		
		{\includegraphics[width=0.34\textwidth]{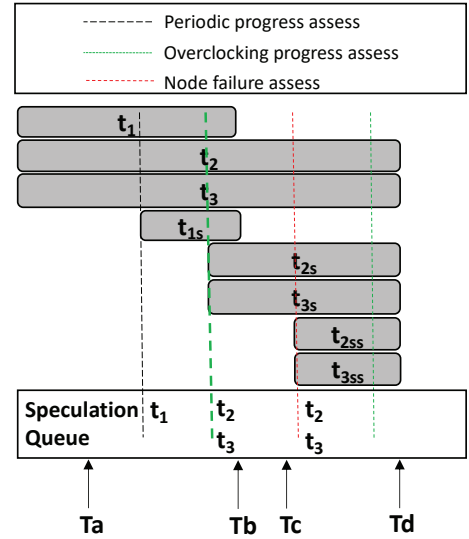}}
		\caption{An example of collective speculation} 
		\vspace{-0.5pc}
		\label{fig:coll_spec}
	\end{center}
\end{figure}

The process of collective speculation is shown by an example in Figure~\ref{fig:coll_spec}.
Here we set {\ttfamily COLL\_INIT\_NUM} to $1$ and {\ttfamily COLL\_MULTIPLY} to $2$.
At time $T_a$, through a progress assessment, 
the node that contains Task $t_1$, $t_2$ and $t_3$ is found to be slow and a speculative copy of $t_{1s}$ is launched.
Then we employ a very small duration for periodic progress checking.
If we find that $t_{1s}$ has higher progress rate than $t_1$, 
we then re-attempt $t_2$ and $t_3$ with a speculative copy. At $T_b$, $t_{1s}$ finishes
and kills the original attempt $t_1$. At $T_c$, the node that contains
speculative tasks $t_{2s}$ and $t_{3s}$ fails,
and is detected promptly by our failure assessment. 
Two further speculative copies are launched for $t_2$ and $t_3$, and
the execution completes at $T_d$. 

To facilitate the dependency-aware speculation, 
we also make speculative copies for completed tasks. 
Upon positive result of failure assessment or 
two consecutive intermediate data fetch failures,
we launch a speculative copy for completed tasks.
Our speculation for completed tasks works in the same way as the speculation for 
incomplete tasks, included to leverage the collective speculation scheme. 
However, when the speculative task completes, 
we do not discard its output. Instead we keep both the original and the speculative outputs 
until successful completion of the job. 

\subsection{Speculative Rollback}
In the existing YARN's speculation, the speculator launches a speculative task
from scratch, the same as YARN's fail-over mechanism where a failed task is rescheduled with a new attempt.
We find that this is not efficient if the original task is delayed because of transient faults,
such as disk I/O exception or packet loss, and the compute node for the task is still available. 
It can be more efficient to launch a task attempt on the original node and start from a previous execution point.
Thus, we design a speculative rollback mechanism within binocular speculation. 
It leverages both YARN's task failure report and our neighborhood assessment.
When a task is reported as slow or failed, 
two attempts of the task are placed on its original node and a new node. 
The speculative copy on the original node will pick up the reserved task progress
and start from there. The other copy will work as ordinary speculative task on another node.
We do not use a heavy-weight remote checkpointing
mechanism~\cite{raft:icde2011} because 
it incurs a lot of additional I/O overheads and would be too costly for short-lived map tasks.

The rollback mechanism provides another advantage because 
multiple attempts of the same task compete for completion. 
To achieve lightweight logs, we keep only a limited set of task information 
that is sufficient for resuming the execution of a map task.
The log includes the {\em spill path} and {\em offset} of the input split for a map task.
When the recovered map task is launched,
it reclaims the completed work done by the previous attempt and rolls directly back 
to the previous offset of input split to start processing from the offset.
In order to avoid making wasteful speculative attempts, 
we also check the status of the previous node. 
If the node is not slow or failed, the rollback speculative task will be scheduled. 
Otherwise, an additional speculation is not allowed. 
Like the original YARN, we will try the speculative attempt on a fast node.
The rollback mechanism is integrated with the other two techniques of binocular speculation, 
validated through sample programs such as TeraValidate from Terasort.

\section{Performance Evaluation}
\label{sec:experiment}

\subsection{Experimental Environment}

All our experiments are conducted on a cluster of 21 server nodes
that are connected through 1 Gigabit Ethernet. Each machine is equipped with four 2.67 GHZ 
hex-core Intel Xeon X5650 CPUs, 24 GB memory and one 500 GB hard disk.
We use YARN 2.7.1 as the code base with JDK 1.7. One
node of the cluster is dedicated to run the ResourceManager of YARN and the NameNode of HDFS. 

\begin{figure*}[!thb]
	\vspace{-0.5pc}
	\begin{center}
		\subfigure[Overall Performance\label{fig:overall}]
		{\includegraphics[width=0.32\textwidth,height=1.35in]{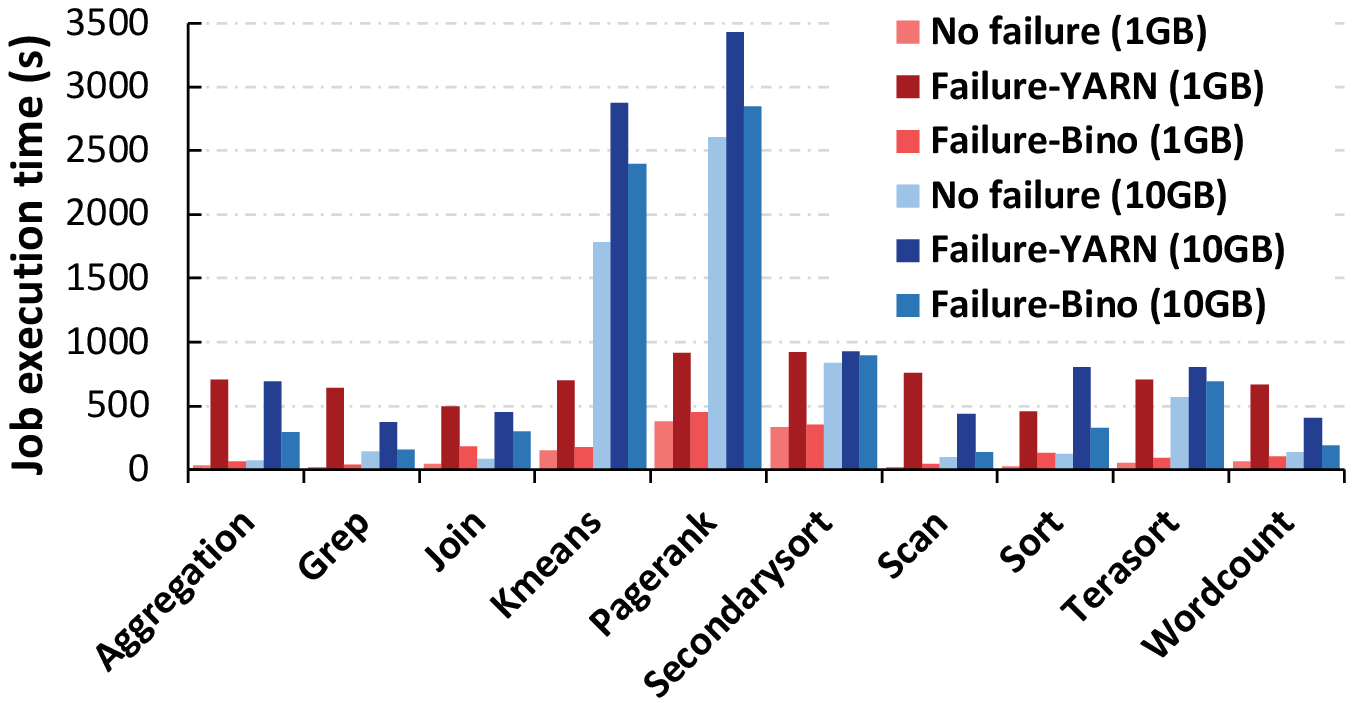}}
		\hspace{0.2em}
		\subfigure[Dependency Awareness\label{fig:retro-spec}]
		{\includegraphics[width=0.32\textwidth]{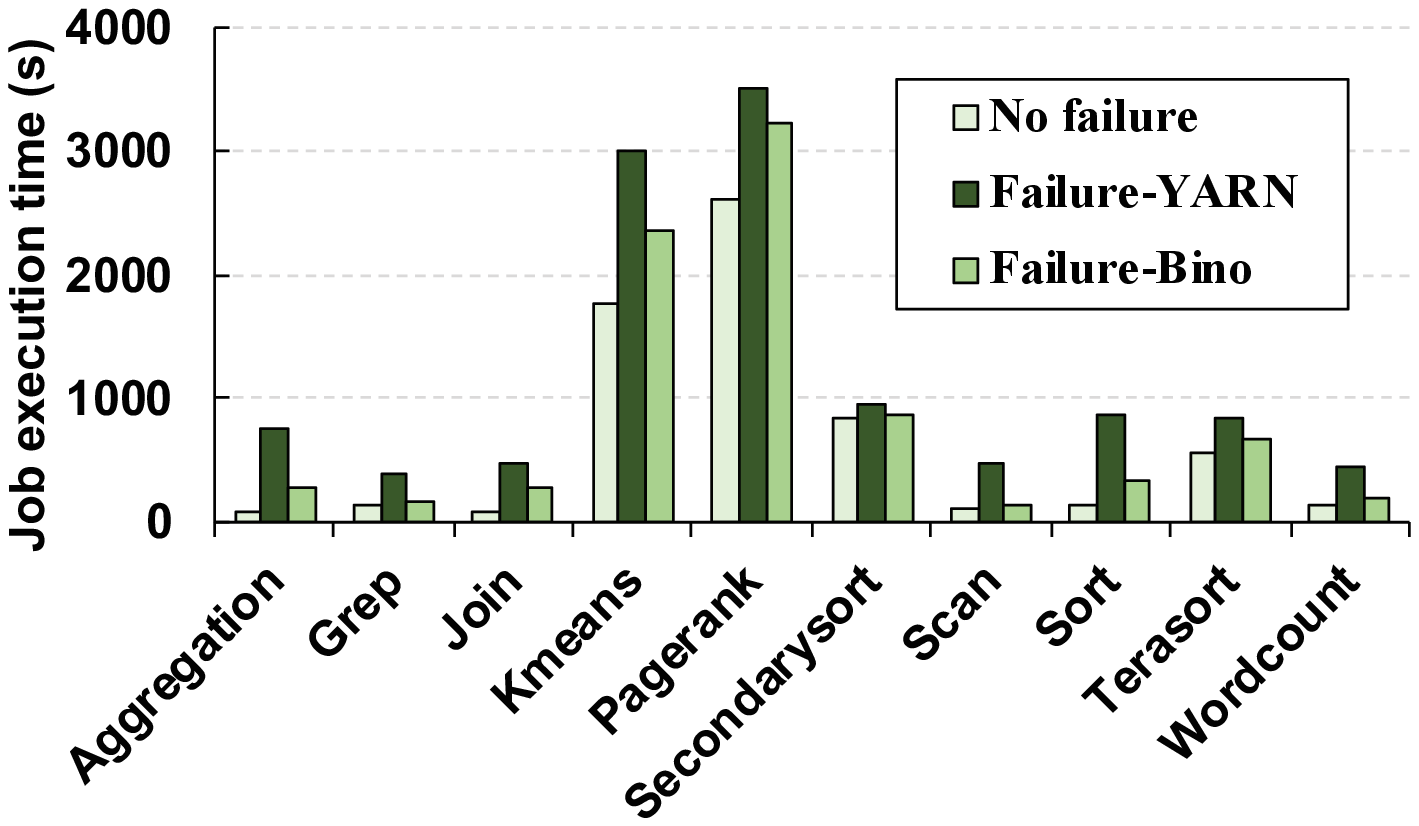}}
		\hspace{0.2em}
		\subfigure[Widening Speculation Scope\label{fig:inter-spec}]
		{\includegraphics[width=0.32\textwidth]{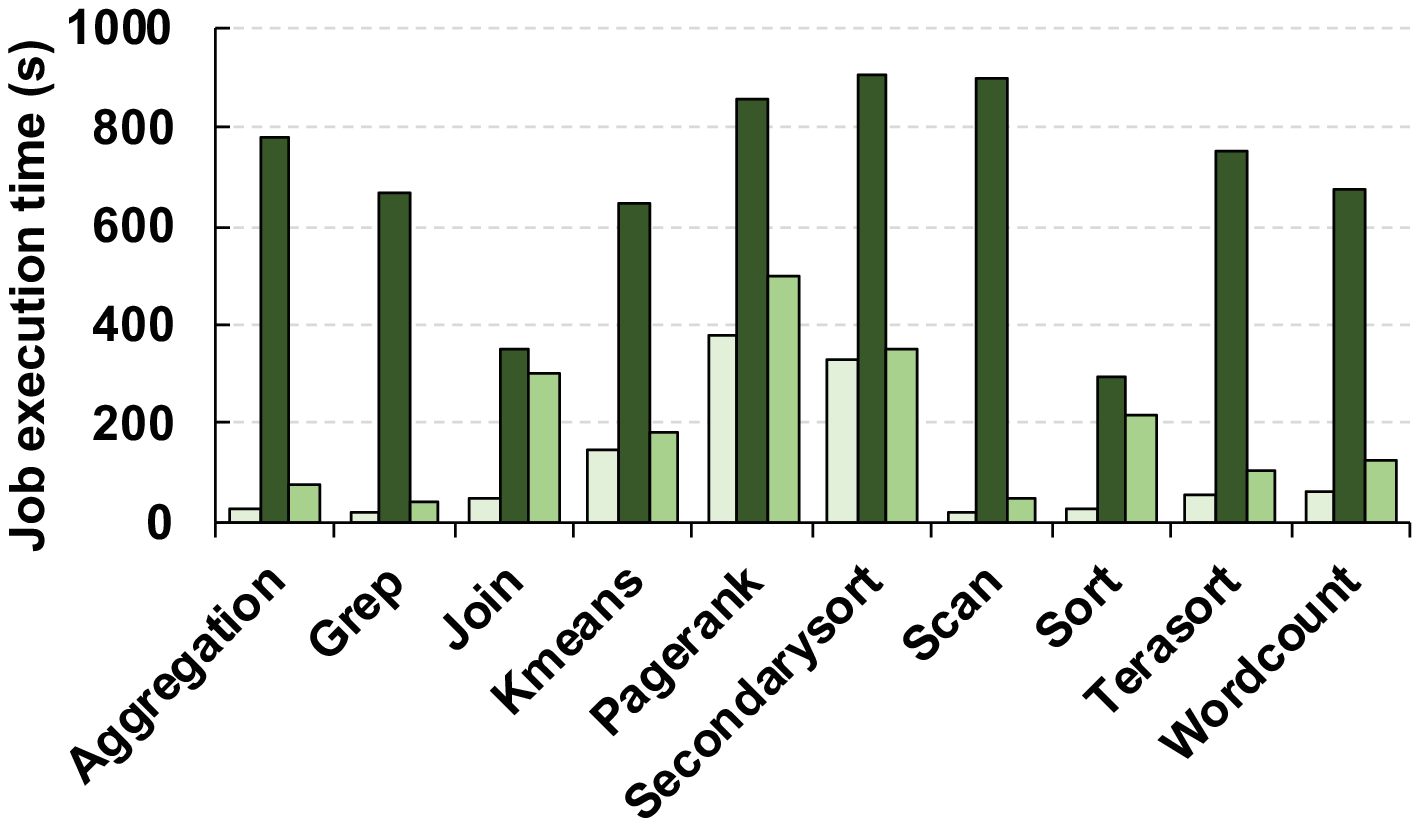}}
		\vspace{-0.5pc}
		\caption{Overall Benefits of Binocular Speculation}
		\vspace{-0.5pc}
		\label{fig:comparison-1g}
	\end{center}
\end{figure*}

\emph{Benchmarks:} 
We have selected a representative set of MapReduce benchmarks,
including {\em Terasort}, {\em Wordcount}, {\em Secondarysort} and {\em Grep} from
YARN's built-in suite and {\em Aggregation}, {\em Join}, {\em Kmeans}, {\em Pagerank},
{\em Scan} and {\em Sort} from the well-known HiBench suite~\cite{huang2010hibench}.
Unless specified, we report the results as the average of the suite of the
benchmarks.

\emph{Evaluation Metrics:} 
 To emulate temporary system faults, we introduce delays in the progress of MapReduce tasks. 
To emulate node failures, we disconnect the targeted compute nodes. 
To measure the efficiency of fault recovery, we compare the job execution time.
In addition, we measure the 
{\em average job slowdown} as the ratio between the job execution time 
upon failures and that without failures. 
To evaluate the gracefulness of binocular speculation in handling stressful
scenarios, we run jobs in a heavily loaded MapReduce cluster and report the distribution
of job execution time as PDF (probability density function) and CDF (cumulative density function) distributions. 

We compare our design of binocular speculation against the original YARN,
which uses the LATE scheduler~\cite{late:osdi08} as its default speculator.
We denote the original speculation as {\bf YARN} and binocular speculation as {\bf Bino}.

\subsection{Overall Benefits of Binocular Speculation}
We conduct experiments to examine the overall benefits of binocular
speculation. We first measure its job execution time with different
speculation policies, then evaluate if binocular speculation can mitigate the symptoms of dependency-oblivious speculation 
and scope-limited speculation.

\subsubsection{Job Execution Time}
Fig.~\ref{fig:overall} shows the job execution time for the two speculation
policies with and without node failures.  
For this experiment, we report the average result from 10 test runs.
For each run, we introduce a node failure at various points of job execution,
i.e., from 10\% to 100\% of map progress.
On average, Bino achieves 7.3x improvement for 1 GB jobs and 1.9x for 10 GB jobs compared to the original YARN.
Some applications, e.g., Aggregation, are very sensitive to node failures and
can experience a job slowdown of more than 20x.
Bino is especially beneficial to these applications due to its speculation coverage and short timeouts.

\subsubsection{Dependency-oblivious speculation}
To validate the mitigation of dependency-oblivious speculation, 
we measure the job execution time only when we observe the loss of intermediate data.
These tests used 10 GB of input data, and the measurements were collected when
there is at least one fetch failure of MOF but no map task failure in order to exclude
the effect of scope-limited speculation. 
Fig.~\ref{fig:retro-spec} shows both YARN and Bino's increments of job execution time upon
the task failure. 
YARN's default speculation is unaware of the dependency of lost data and causes
the jobs to run much longer, on average a performance slowdown of 
4.0x compared to the case without failure. 

In contrast, with dependency awareness, binocular speculation pinpoints
the corresponding map task for the lost intermediate data and launches a speculative task for timely recomputation.
Thus compared to the default YARN speculation, binocular speculation provides
an improvement of 2.0x average to these benchmarks.

\subsubsection{Scope-limited speculation}
We also test scope-limited speculation with 1 GB of input data and only
adopt the result that is affected solely by the scope-limited speculation,
i.e., there are failed map tasks that are not timely speculated
but no MOF is lost.

Fig.~\ref{fig:inter-spec} shows Bino's benefits in restoring the
scope-limited speculation from inactivity and recover the tasks effectively.
Since the jobs are relatively small here, a node failure can cause significant performance degradation to the benchmarks. .
Binocular speculation in this case provides a much bigger improvement (an
average of 6.8x) because it can quickly detect the failure-related stragglers.

\subsection{Impact to Task Skew and Job Slowdown}

A task or node failure can have very different impact on different tasks in a
job because of their proximity or dependency. A good speculation shall 
overcome the disruptive impact of system faults, and smoothen the
execution times of different tasks.
We compare the two speculation policies by measuring 
the distribution of task execution times and the overall job slowdown upon a
node failure.

\begin{figure}[tbh]
	\centering
	\vspace{-0.5pc}
	\includegraphics[width=0.38\textwidth]{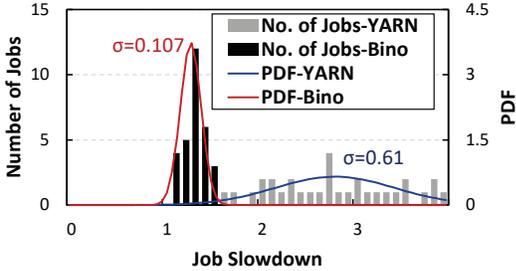}
	\caption{Performance variance}
	\vspace{-0.5pc}
	\label{fig:variation-test}
\end{figure}

Fig.~\ref{fig:variation-test} shows the PDF distribution of the average job
slowdown for the benchmarks under two different speculation policies, as well as their standard variations. 

YARN default speculation leads to a much wider distribution of job
slowdown with an average around 2.8. Binocular speculation significantly
reduces the average slowdown and decreases the variance $\sigma$ from 0.61 to 0.107.

\subsection{System Efficiency Under Stress}
\label{sec:heavy-test}
\begin{figure}[tbh]
	\centering
	\includegraphics[width=.38\textwidth]{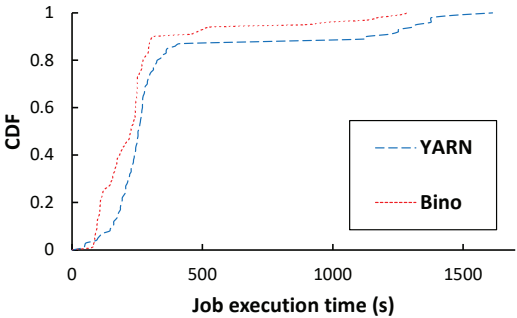}
	\caption{Performance under stress}
	\vspace{-0.5pc}
	\label{fig:overall-test1}
\end{figure}

We evaluate how binocular speculation performs when the 
system is under a heavy load of many concurrent jobs.
Because many jobs are competing for shared resources, more tasks 
will experience significantly slower progress~\cite{delay:eurosys10,wang:icac13}.
We run Terasort, Wordcount, Secondarysort and Grep jobs with different input sizes.
We have followed the PACMan work~\cite{ananthanarayanan2012pacman} to set the size of jobs.
85\% of jobs have 1 GB input data, 8\% with 10 GB, and 5\% with 50 GB and the rest with 100 GB.
We let the job arrive at random times following a Poisson distribution.
We then inject task failures, node crashes and network delays.
Fig.~\ref{fig:overall-test1} measures the job execution times and plots their CDF distribution.
Binocular speculation leads to significant improvement for MapReduce jobs. A greater improvement can be observed for small jobs.
Big jobs can still benefit from binocular speculation, but relatively less due to the bigger tasks.
Both YARN and Bino have tails of long-running jobs that are due to the long turnaround time of big jobs.
Our results suggest that binocular speculation can also handle stressful situations for heavy workloads on production systems.
On average, it decreases the job execution time of all jobs in our synthetic workload by 30\%.

\subsection{Analysis and Tuning of Component Techniques}
We evaluate the performance of individual techniques in binocular speculation
and tune some of the key configuration parameters. We describe our evaluation
and analysis in the rest of this section.

\subsubsection{Neighborhood Glance}
\begin{figure*}[thb]
	\vspace{-0.5pc}
	\begin{center}
		\subfigure[Performance dissection.\label{fig:tuning-assess}]
		{\includegraphics[width=0.32\textwidth]{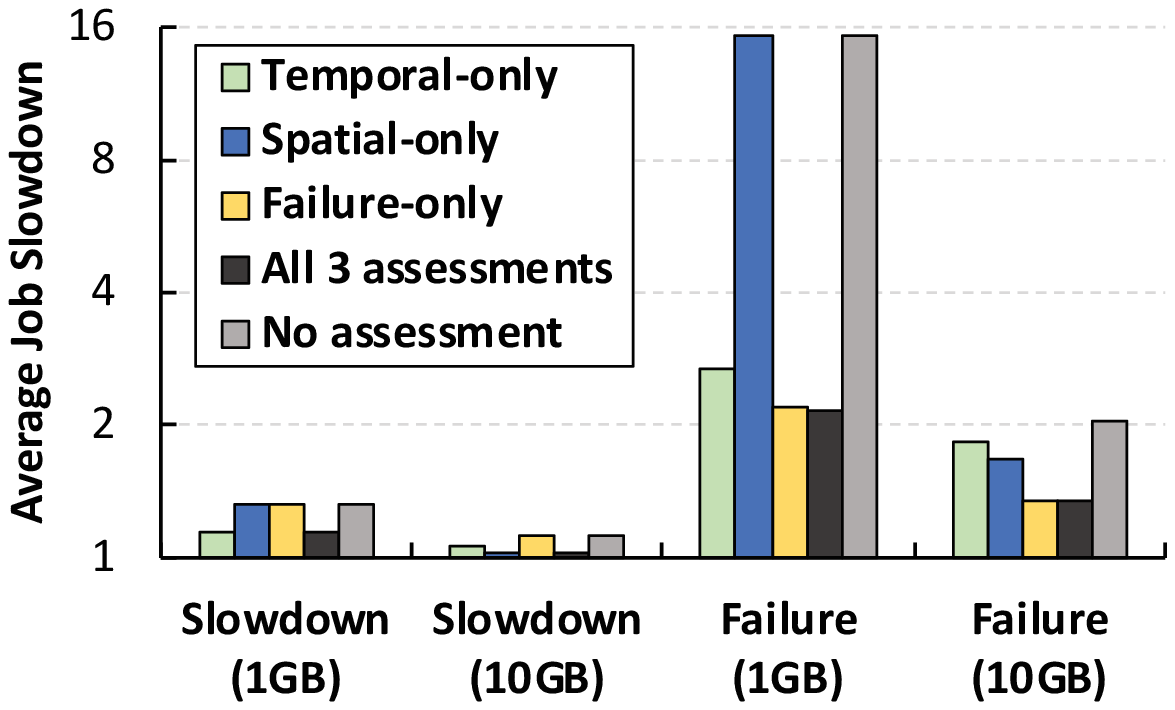}}
		\hspace{0.2em}
		\subfigure[Correctness of failure assessment.\label{fig:fail-aware}]
		{\includegraphics[width=0.32\textwidth]{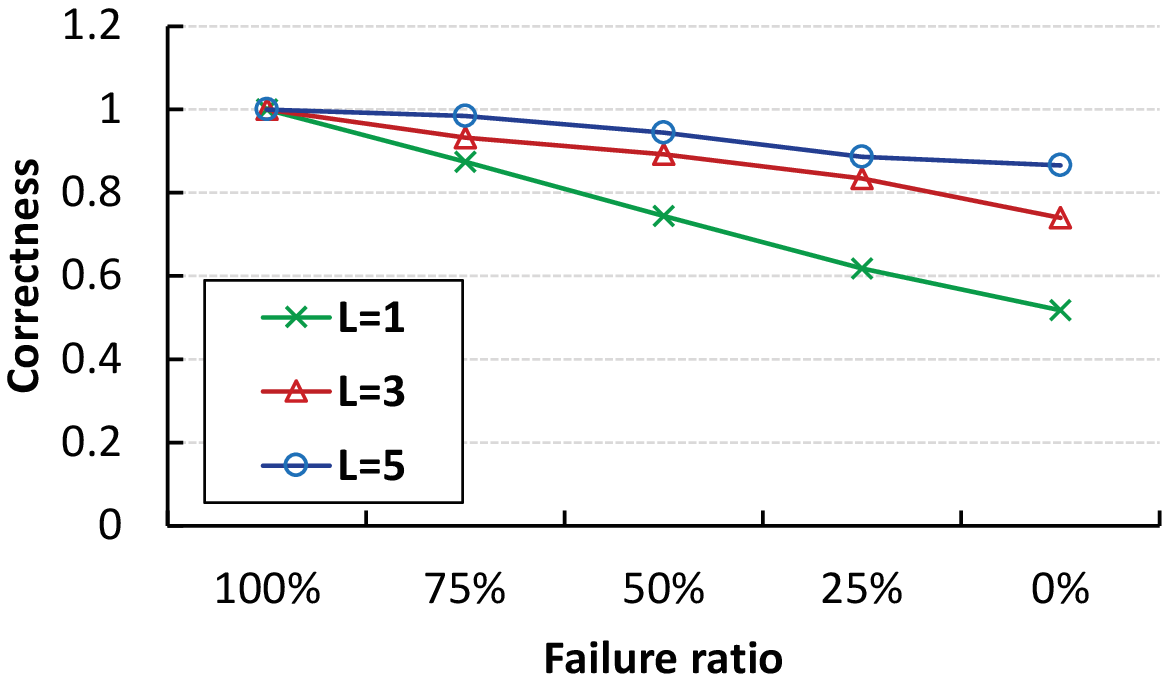}}
		\hspace{0.2em}
		\subfigure[Tuning neighborhood size.\label{fig:tuning-neigh}]
		{\includegraphics[width=0.32\textwidth]{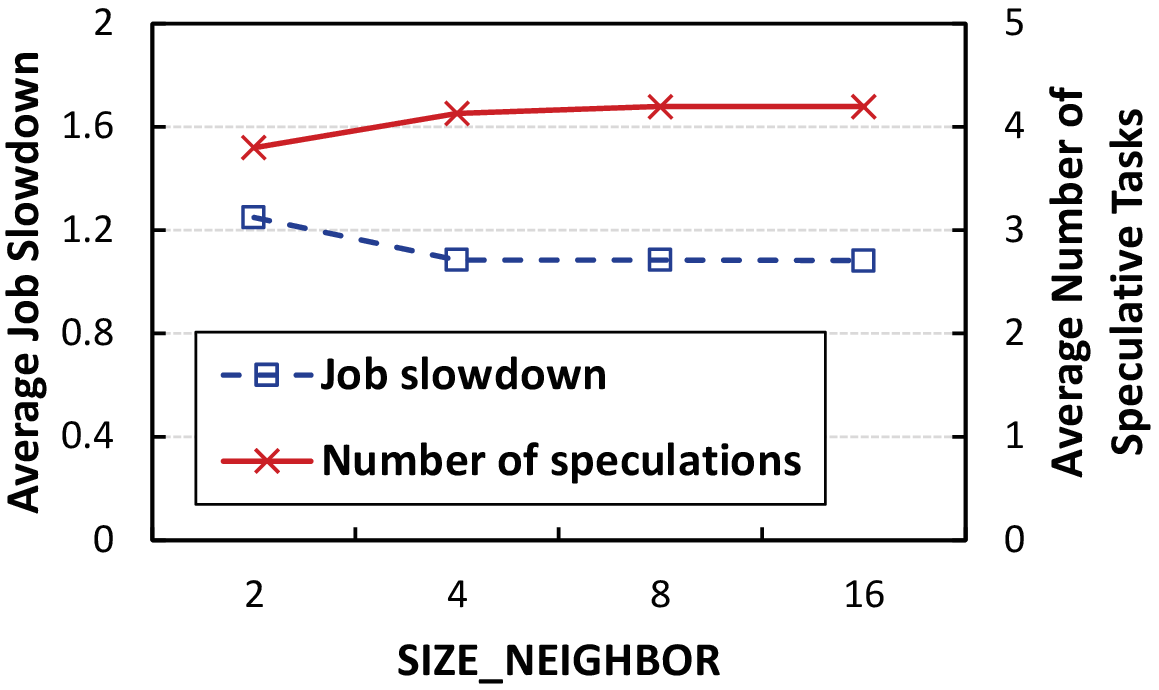}}
		\vspace{-0.5pc}
		\caption{Understanding neighborhood glance}
		\vspace{-0.5pc}
		\label{fig:tuning-neigh-glance}
	\end{center}
\end{figure*}

We evaluate the performance of neighborhood glance.
For this experiment, we selectively enable only one or more of the assessment policies,
against a node delay or failure with different job sizes.
As shown in Fig.~\ref{fig:tuning-assess}, 
different assessment has distinct effects on mitigating the job slowdown.
When enabling only the temporal assessment,
binocular speculation can significantly mitigate the performance degradation for small jobs,
achieving an average of 5.7x improvement for node failures.
In addition, our experiments show that enabling failure assessment provides 
comparable or even larger improvement than detecting temporal progress changes.
Moreover, spatial assessment does not help node failures in small jobs but
it can mitigate node failures in larger jobs better than temporal assessment,
e.g. an average of 22.1\% improvement for node failures.
The results also show that node slowdown can 
cause smaller but notable performance degradation,
and our neighborhood glance has similar effect on the recovery.

We further evaluate the failure assessment by tuning a key parameter $L$
(window size) used in the temporal window-based assessment, 
which is the number of prior unresponsive scenarios that Bino takes into account.
We examine the correctness of our node failure assessment in a heterogeneous environment 
where both node failures and network delays are common.
Thus, in each test, we inject a number of node failures and/or network delays.
The number of failures and delays are following a {\em failure ratio}, varying from 0\% to 100\%.
The duration of delays is randomly generated according to a Poisson Distribution.

As shown in Fig.~\ref{fig:fail-aware}, there are two major trends.
Firstly, the larger the failure ratio is, the more accurate our failure assessment is.
Secondly, to set $L$ higher also leads to higher accuracy.
Furthermore, we investigate the impact of {\ttfamily SIZE\_NEIGHBOR}, 
i.e., the number of nodes in a neighborhood.
We manually slow down a compute node during the job execution, and then
measure the job slowdown and the number of speculative tasks for the job to complete.

As shown in Fig.~\ref{fig:tuning-neigh}, {\ttfamily SIZE\_NEIGHBOR} does not have a very large impact
on job slowdown or the number of speculative tasks.
However, a neighborhood with only two nodes has smaller performance improvement 
due to its limited capacity for spatial progress assessment.

This implies that we can keep {\ttfamily SIZE\_NEIGHBOR} small to minimize
cross traffic I/O overheads so long as enough progress variation can be
detected within the neighborhood.

\begin{figure}[tbh]
	\centering
	\vspace{-0.5pc}
	\includegraphics[width=0.36\textwidth]{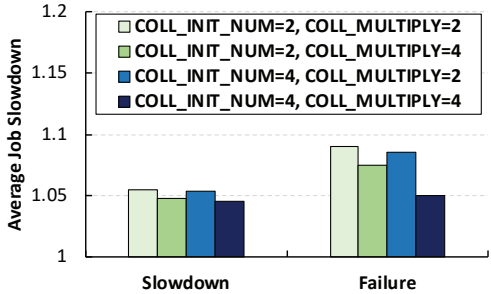}
	\caption{Tuning collective speculation}
	\vspace{-0.5pc}
	\label{fig:tuning-coll}
\end{figure}

\subsubsection{Collective Speculation}

{\ttfamily COLL\_INIT\_NUM} and {\ttfamily COLL\_MULTIPLY} are two critical parameters
in the collective speculation.
 
We tune those parameters when a node is delayed in progress or failed.
The results are shown in Fig.~\ref{fig:tuning-coll}.
Overall, increasing {\ttfamily COLL\_MULTIPLY} has a bigger impact
to the job performance for both node delays and failures. 
Increasing {\ttfamily COLL\_INIT\_NUM} reduces the average job slowdown but has a smaller impact.
However, launching more speculative tasks aggressively can 
consume resources very quickly. 
It can be tolerable when the system workload is light, but can be very
disruptive to all users and jobs on a shared system that is heavily loaded.

\subsubsection{Speculative Rollback}

\begin{figure}[tbh]
	\vspace{-0.5pc}
	\centering
	\includegraphics[width=0.38\textwidth]{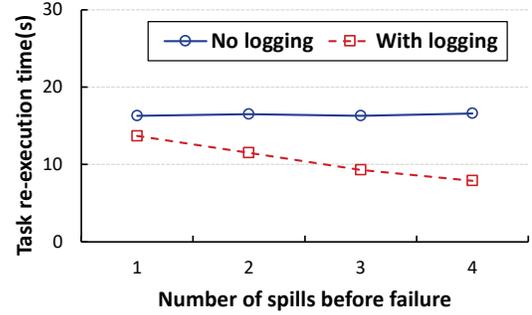}
	\caption{Benefits of speculative rollback}
	\vspace{-0.5pc}
	\label{fig:maploggin-task}
\end{figure}

We conduct experiments to demonstrate the benefits of speculative rollback. We inject map task failures by incurring a disk write exception to a single map task.
For each job we inject only one failure but at a different progress point.
The progress point of a task is indicated by the number of spills it has generated before the failure.
Fig.~\ref{fig:maploggin-task} shows the performance improvement from the speculative rollback 
to a new attempt of the failed task.
We can see that the actual performance gain depends on the amount of task progress that has reached before failures.
When there is more task progress, the speculative rollback recovery is faster. 
For instance, re-execution for a failure after 4 spills takes 73\% shorter time than the one after 1 spill,
effectively preserving map task progress and speeding up job recovery.

\section{Related Work}
\label{sec:related}
\subsection{MapReduce Speculation}
Speculation mechanism had been actively studied with a variety of 
viewpoints~\cite{late:osdi08,osdi:mantri,nsdi14grass,ganesh:clones}.
But compared to our work, they either overlooked the particular behaviors of failure-related
stragglers and/or adopted a mitigation strategy that is inefficient.
To describe a few, LATE~\cite{late:osdi08} scheduler took node heterogeneity into account.
But its speculator scope was limited and it used the suboptimal serial speculation. 
Mantri~\cite{osdi:mantri} searched for the causes of stragglers 
and identified in part the impact of failure-related stragglers.
Moreover, it made many interesting observations.
For example, the failure-related stragglers are very often localized on a few bad machines
but those machines are usually scattered apart across the cluster.
However, it also had a limited scope of speculation
and only used the costly replication of intermediate data to avoid data recomputation during failures.
GRASS~\cite{nsdi14grass} improved speculation
only for the  
approximation jobs
using two distinct scheduling strategies, 
a.k.a. Greedy Speculative and Resource Aware Speculative scheduling.
But neither of the two strategies addressed failure recovery.

DOLLY~\cite{ganesh:clones} used undifferentiated task cloning to mitigate stragglers in small jobs. Although such design could be helpful for solving the 
performance breakdown of node failures, it had an obvious downside that cloning every task would incur a lot more computation and network I/O than making only necessary speculation for a few stragglers. This is especially true for a shared MapReduce cluster that is already heavily loaded as discussed in~\cite{jian:sigmetrics12, wang:icac13}. 
Without handling failures respectively, relying on such exhaustive speculation for
mitigating failure-related stragglers is not practical.
To the best of our knowledge, our work is the first one to examine the
failure-related stragglers in MapReduce, and accordingly, presents an effective and efficient mitigation solution.

\subsection{MapReduce Failure Recovery}

Besides speculation, MapReduce failure resiliency has also gained much attention.
Since failures are gradually becoming the norm in large-scale systems, 
the recovery efficiency is equally important.
Many recent studies have been working towards efficient failure recovery for MapReduce 
systems~\cite{raft:icde2011,dinu:hpdc12,wang:ipdps15}. 
RAFTing MapReduce~\cite{raft:icde2011} for preserving the computation of map tasks
and replicating the MOFs to reduce side. This design avoided the 
re-computation of map tasks on the failed node, but it required
the pre-assignment of reduce tasks and incurred much additional network overheads.

Dinu {\em et al.} \cite{dinu:hpdc12} took
an experimental analysis on the node failure in MapReduce.
They revealed an issue called {\em delayed speculative execution} during node failure.
Moreover, they found that the failure of the node containing reduce tasks
can infect other healthy tasks and nodes, causing drastic performance degradation.
Wang {\em et al.}~\cite{wang:ipdps15} revealed a similar issue, 
which is referred to as {\em failure amplification}.
But both works failed to recognize the inherent cause, i.e., 
the issue of dependency oblivious speculation so neither of them provided
an efficient solution for this issue. Both works did not look into failures occurring at the map phase.
Our work is orthogonal to these prior studies by addressing speculation myopia and
its impact on fault recovery in MapReduce.

\section{Conclusion}

\label{sec:conclusion}

In this paper, we have examined the dichotomy of MapReduce caused by its two-phase
execution model and then used the next-generation Hadoop framework, i.e., YARN,
to examine the role of MapReduce speculation for failure recovery.
We reveal that MapReduce dichotomy leads to a problem called speculation
myopia, which has impact on MapReduce fault recovery upon temporary system faults and/or 
node failures. Speculation myopia often manifests itself in two main symptoms:
dependency-oblivious speculation and scope-limited speculation.
We have designed and implemented a new speculation scheme
called binocular speculation to address speculation myopia and its symptoms. Binocular speculation is designed with three constituent techniques including 
neighborhood glance, collective speculation and speculative rollback.
We have conducted a substantial set of experiments to evaluate the benefits of
binocular speculation for performance breakdown and variations on 
MapReduce systems. Our experimental results demonstrate 
that binocular speculation can heal all symptoms of speculation myopia 
and deliver fast fault recovery compared to the existing speculation mechanism 
in YARN MapReduce. 

\section*{Acknowledgment}

This work is supported in part by the National Science Foundation awards 1561041, 1564647, and 1744336.


\begin{thebibliography}{10}

\bibitem{yarn_url}
Apache hadoop nextgen mapreduce (yarn).
\newblock
  \url{http://hadoop.apache.org/docs/r2.3.0/hadoop-yarn/hadoop-yarn-site/YARN.html}.

\bibitem{ganesh:clones}
Ganesh Ananthanarayanan, Ali Ghodsi, Scott Shenker, and Ion Stoica.
\newblock Effective straggler mitigation: Attack of the clones.
\newblock NSDI '13.

\bibitem{ananthanarayanan2012pacman}
Ganesh Ananthanarayanan, Ali Ghodsi, Andrew Wang, Dhruba Borthakur, Srikanth
  Kandula, Scott Shenker, and Ion Stoica.
\newblock Pacman: coordinated memory caching for parallel jobs.
\newblock NSDI '12.

\bibitem{nsdi14grass}
Ganesh Ananthanarayanan, Michael Chien-Chun Hung, Xiaoqi Ren, Ion Stoica, Adam
  Wierman, and Minlan Yu.
\newblock Grass: Trimming stragglers in approximation analytics.
\newblock In {\em 11th USENIX Symposium on Networked Systems Design and
  Implementation (NSDI 14)}, pages 289--302, Seattle, WA, April 2014. USENIX
  Association.

\bibitem{osdi:mantri}
Ganesh Ananthanarayanan, Srikanth Kandula, Albert Greenberg, Ion Stoica, Yi~Lu,
  Bikas Saha, and Edward Harris.
\newblock Reining in the outliers in map-reduce clusters using mantri.
\newblock OSDI'10.

\bibitem{appuswamy2013scale}
Raja Appuswamy, Christos Gkantsidis, Dushyanth Narayanan, Orion Hodson, and
  Antony Rowstron.
\newblock Scale-up vs scale-out for hadoop: Time to rethink?
\newblock SOCC '13.

\bibitem{chen2012interactive}
Yanpei Chen, Sara Alspaugh, and Randy Katz.
\newblock Interactive analytical processing in big data systems: A
  cross-industry study of mapreduce workloads.
\newblock {\em Proceedings of the VLDB Endowment}, 5(12), 2012.

\bibitem{dean2006experiences}
Jeffrey Dean.
\newblock Experiences with mapreduce, an abstraction for large-scale
  computation.
\newblock In {\em PACT}, volume~6, pages 1--1, 2006.

\bibitem{osdi/googlemr04}
Jeffrey Dean and Sanjay Ghemawat.
\newblock Mapreduce: Simplified data processing on large clusters.
\newblock OSDI '04.

\bibitem{dinu:hpdc12}
Florin Dinu and T.S.~Eugene Ng.
\newblock Understanding the effects and implications of compute node related
  failures in hadoop.
\newblock HPDC '12.

\bibitem{dong:sc09}
Xiangyu Dong, Naveen Muralimanohar, Norm Jouppi, Richard Kaufmann, and Yuan
  Xie.
\newblock Leveraging 3d pcram technologies to reduce checkpoint overhead for
  future exascale systems.
\newblock In {\em Proceedings of the Conference on High Performance Computing
  Networking, Storage and Analysis}, SC '09, pages 57:1--57:12, New York, NY,
  USA, 2009. ACM.

\bibitem{DRFsched}
Ali Ghodsi, Matei Zaharia, Benjamin Hindman, Andy Konwinski, Scott Shenker, and
  Ion Stoica.
\newblock Dominant resource fairness: fair allocation of multiple resource
  types.
\newblock In {\em Proceedings of the 8th USENIX conference on Networked systems
  design and implementation}, NSDI'11, pages 24--24, Berkeley, CA, USA, 2011.
  USENIX Association.

\bibitem{huang2010hibench}
Shengsheng Huang, Jie Huang, Jinquan Dai, Tao Xie, and Bo~Huang.
\newblock The hibench benchmark suite: Characterization of the mapreduce-based
  data analysis.
\newblock ICDEW '10.

\bibitem{isard2007dryad}
Michael Isard, Mihai Budiu, Yuan Yu, Andrew Birrell, and Dennis Fetterly.
\newblock Dryad: distributed data-parallel programs from sequential building
  blocks.
\newblock In {\em SIGOPS}, 2007.

\bibitem{4-kharbascombining}
K.~Kharbas, D.~Fiala, F.~Mueller, K.~Ferreira, and C.~Engelmann.
\newblock {Combining Partial Redundancy and Checkpointing for HPC}.
\newblock In {\em {International Conference on Distributed Computing Systems}},
  2012.

\bibitem{moody:sc10}
Adam Moody, Greg Bronevetsky, Kathryn Mohror, and Bronis R.~de Supinski.
\newblock Design, modeling, and evaluation of a scalable multi-level
  checkpointing system.
\newblock In {\em Proceedings of the 2010 ACM/IEEE International Conference for
  High Performance Computing, Networking, Storage and Analysis}, SC '10, pages
  1--11, Washington, DC, USA, 2010. IEEE Computer Society.

\bibitem{raft:icde2011}
Jorge-Arnulfo Quiane-Ruiz, Christoph Pinkel, Jorg Schad, and Jens Dittrich.
\newblock Rafting mapreduce: Fast recovery on the raft.
\newblock ICDE '11.

\bibitem{reiss2012heterogeneity}
Charles Reiss, Alexey Tumanov, Gregory~R Ganger, Randy~H Katz, and Michael~A
  Kozuch.
\newblock Heterogeneity and dynamicity of clouds at scale: Google trace
  analysis.
\newblock SOCC '12.

\bibitem{schroeder:dsn06}
Bianca Schroeder and Garth~A. Gibson.
\newblock A large-scale study of failures in high-performance computing
  systems.
\newblock In {\em Proceedings of the International Conference on Dependable
  Systems and Networks}, DSN '06, pages 249--258, Washington, DC, USA, 2006.
  IEEE Computer Society.

\bibitem{jian:sigmetrics12}
Jian Tan, Xiaoqiao Meng, and Li~Zhang.
\newblock Delay tails in mapreduce scheduling.
\newblock SIGMETRICS '12.

\bibitem{tan:sigmetrics14}
Jian Tan, Yandong Wang, Weikuan Yu, and Li~Zhang.
\newblock Non-work-conserving effects in mapreduce: Diffusion limit and
  criticality.
\newblock In {\em Proceedings of the 14th ACM SIGMETRICS/PERFORMANCE Joint
  International Conference on Measurement and Modeling of Computer Systems},
  SIGMETRICS '14, June 2014.

\bibitem{3-wang2010hybrid}
C.~Wang, F.~Mueller, C.~Engelmann, and S.L. Scott.
\newblock {Hybrid Checkpointing for MPI Jobs in HPC Environments}.
\newblock In {\em {16th International Conference on Parallel and Distributed
  Systems}}, 2010.

\bibitem{wang:ipdps15}
Yandong Wang, Huansong Fu, and Weikuan Yu.
\newblock Cracking down mapreduce failure amplification through analytics
  logging and migration.
\newblock IPDPS '15.

\bibitem{wang:icac13}
Yandong Wang, Jian Tan, Weikuan Yu, Xiaoqiao Meng, and Li~Zhang.
\newblock Preemptive reducetask scheduling for fair and fast job completion.
\newblock In {\em Proceedings of the 10th International Conference on Autonomic
  Computing}, ICAC'13, June 2013.

\bibitem{sigmod:hadoop:dbms}
Yu~Xu, Pekka Kostamaa, and Like Gao.
\newblock Integrating hadoop and parallel dbms.
\newblock In {\em Proceedings of the 2010 ACM SIGMOD International Conference
  on Management of data}, pages 969--974. ACM, 2010.

\bibitem{delay:eurosys10}
Matei Zaharia, Dhruba Borthakur, Joydeep Sen~Sarma, Khaled Elmeleegy, Scott
  Shenker, and Ion Stoica.
\newblock Delay scheduling: A simple technique for achieving locality and
  fairness in cluster scheduling.
\newblock EuroSys '10.

\bibitem{late:osdi}
Matei Zaharia, Andy Konwinski, Anthony~D. Joseph, Randy Katz, and Ion Stoica.
\newblock Improving mapreduce performance in heterogeneous environments.
\newblock OSDI'08.

\bibitem{late:osdi08}
Matei Zaharia, Andy Konwinski, Anthony~D. Joseph, Randy Katz, and Ion Stoica.
\newblock Improving mapreduce performance in heterogeneous environments.
\newblock In {\em Proceedings of the 8th USENIX conference on Operating systems
  design and implementation}, OSDI'08, pages 29--42, Berkeley, CA, USA, 2008.
  USENIX Association.

\end{thebibliography}
\end{document}